\documentstyle[12pt]{article}

\begin{document}

\title{Branching Ratio and direct CP Asymmetry
for $B \rightarrow K^{*} \gamma$ Decay in MSSM}
\author{Marina-Aura Dariescu
and Ciprian Dariescu\thanks{On leave of
absence from Department of Theoretical Physics,
{\it Al. I. Cuza} University, Bd. Copou no. 11,
6600 Ia\c{s}i, Romania, email address (after May 1):
marina@uaic.ro} \\
{\it Institute of Theoretical Science} \\
{\it University of Oregon, Eugene, OR 97403}}
\date{}

\maketitle

\begin{abstract}
The present paper deals with a next-to-leading order
analysis of the radiative $B \to K^* \gamma$ decay.
Working in the PQCD approach developed
by Szczepaniak {\it et al.}, we compute the   
correction, coming from a single gluon exchange
with the spectator, to the essential form factor.
Since the branching ratio gets much
above the experimental data, although in agreement with
other theoretical models predictions,
we take into consideration the effects
of complex flavour couplings
in the squark sector. Finally, we discuss these
SUSY implications on the branching ratio and
direct CP asymmetry values and impose bounds on the  
squark mixing parameter
$\left( \delta_{23}^d \right)_{LR}$.

\end{abstract}

\newpage

After the Cabibbo-favoured $b \to s \gamma$ mode
was first reported, in 1993,
by CLEO II [1] and updated in 1995 [2],
the exclusive radiative decays,
$B \to K^* \gamma$ and $B \to \rho \gamma$,
as well as the inclusive ones, $B \to X_{s(d)} \gamma$,
have become main targets for both
experimental and theoretical investigations.
The exclusive modes, which are easier to be
experimentally investigated [3, 4, 5],
but less theoretically clear,
have been worked out in different
approaches. For example,
the spin symmetry for heavy quarks combined with
wave function models [6, 7]
or the heavy quark effective theory
when both $b$ and $s$ are heavy [8] have been used.
Also, perturbative QCD (PQCD) formalisms, introduced
for exclusive nonleptonic heavy-to-light transitions,
have been extended to account for the radiative decays. 
Recently, detailed analyses of $B \to K^*
\gamma$ and $B \to \rho \gamma$, in the
next-to-leading order (NLO), with the inclusion
of hard spectator and vertex corrections,
have been performed [9-11] and a
consistent treatment, based on a new factorization
formula, has been proposed [12]. 
Besides an independent determination of
the $| V_{td} / V_{ts}|$ ratio,
the $b \to s \gamma$ decays are suitable for studying the
viability of SUSY extensions of the SM, in view of
flavour changing neutral currents (FCNC) and CP tests,
and for imposing constraints on the supersymmetric
benchmark scenarios [13, 14].

The aim of the present paper is to analyse the
$B \to K^* \gamma$ decay, in the minimal
supersymmetric SM (MSSM) context.
First, at next-to-leading order,
we compute the hard-spectator correction
to the essential form factor. In this respect,
we employ the PQCD approach developed
by Szczepaniak {\it et al.} [15], for decays
dominated by tree diagrams and later extended to
^^ ^^ penguin'' processes [16].
As the branching ratio gets much
above the experimental data, we make use
of the mass insertion method to include, in the
Wilson coefficients $C_{7,8}$, gluino-mediated
FCNC contributions.
Finally, the available data on $Br$ and
direct CP asymmetry
are used to constrain the complex values
of the squark mixing parameter
$\left( \delta_{23}^d \right)_{LR}$.

The effective Hamiltonian which describes
the $B \to K^* \gamma$ radiative decay
is given by [9, 10]
\begin{equation}
H \, = \, \frac{G_{F}}{\sqrt{2}} \, \lambda_p
\left[ C_{7} {\cal O}_{7} 
+ C_1 {\cal O}^p_1 + C_8 {\cal O}_8 \right] \, ,
\end{equation}
where $\lambda_p \equiv  V_{pb} V_{ps}^{*}$, with
$p$ summed over $u$ and $c$, and
$C_1$, $C_{7}$, $C_8$ are the effective
Wilson coefficients at $\mu =m_b$.
The hadronic matrix elements of the
four-fermion operator and of the electromagnetic
and chromagnetic penguin operators
\begin{eqnarray}
{\cal O}_1^p & = &
\left( \bar{s} \gamma_{\mu} (1-\gamma_5 ) p \right)
\left( \bar{p} \gamma^{\mu} (1-\gamma_5 ) b \right)
\nonumber \\*
{\cal O}_{7} & = & \frac{e m_b}{8 \pi^{2}} 
\, \left[
\bar{s} \sigma^{\mu \nu} (1+\gamma_{5}) b
\right] F_{\mu \nu}
\nonumber \\*
{\cal O}_{8} & = & \frac{g_s m_b}{8 \pi^{2}} 
\, \left[
\bar{s} \sigma^{\mu \nu} (1+\gamma_{5}) T^i b
\right] G^i_{\mu \nu} 
\end{eqnarray}
possess a general Lorentz decomposition
\begin{eqnarray}
\langle K^* | \bar{s} \gamma_{\mu} b | \bar{B} \rangle
& = &
\frac{2 i \, V(q^2)}{m_B + m_{K^*}} \, 
\varepsilon_{\mu \nu \alpha \beta}
\epsilon^{* \nu} P_{K^*}^{\alpha} P_B^{\beta} \, ,
\nonumber \\*
\langle K^* | \bar{s} \gamma_{\mu} \gamma_5 
b | \bar{B} \rangle & = &
2 m_{K^*} A_0 (q^2) \frac{(\epsilon^* q)}{q^2} \,
q_{\mu} \nonumber \\*
& + &
A_1 (q^2) (m_B + m_{K^*} )
\left[ \epsilon^*_{\mu} -
\frac{(\epsilon^* q)}{q^2} \, q_{\mu} \right]
\nonumber \\*
& - &
A_2 (q^2) \, \frac{(\epsilon^* q)}{m_B + m_{K^*}} 
\left[ ( P_B + P_{K^*})_{\mu} -
\frac{(m_B^2-m_{K^*}^2)}{q^2} \, q_{\mu} \right] ,
\; \; \; \; \; \;
\\*
\langle K^* | \bar{s} \sigma_{\mu \nu}
q^{\nu} b | \bar{B} \rangle & = &
2 \, T_1(q^2) \, \varepsilon_{\mu \nu \alpha \beta}
\epsilon^{* \nu} P_{K^*}^{\beta} P_{B}^{\alpha} \, ,
\nonumber \\*
\langle K^* | \bar{s} \sigma_{\mu \nu} \gamma_5
q^{\nu} b | \bar{B} \rangle & = &
- \, i \, 
T_2(q^2) \left[ \left( m_B^2 - m_{K^*}^2 \right)
\epsilon^{*}_{\mu}
- ( \epsilon^* q) \left( P_B + P_{K^*} \right)_{\mu}
\right]
\nonumber \\*
& - & i \,
T_3 (q^2) ( \epsilon^* q) \left[ q_{\mu} -
\frac{q^2}{m_B^2 -m_{K^*}^2} \,
\left( P_B + P_{K^*} \right)_{\mu} \right],
\end{eqnarray}
where $q_{\mu}$ is the
the momentum of the photon and $\epsilon^{\nu}$ is the
$K^*$ 4-vector polarization.
In the heavy quark limit, $m_b \gg \Lambda_{QCD}$,
neglecting the corrections of order $1/m_b$
and $\alpha_s$, one has the following relation
among the form factors [9]
\begin{equation}
\frac{m_B}{m_B + m_{K^*}} \, V(0) \, = \,
\frac{m_B + m_{K^*}}{m_B} \, A_1(0) \, = \,
T_1(0) \, = \, T_2(0) \equiv F_{K^*}(0)
\end{equation}
This relation is broken when one includes QCD radiative
corrections coming from vertex renormalization and
hard gluon exchanges with the spectator.
We recommend [9, 10, 12]
for detailed analyses of both
factorizable and nonfactorizable vertex and hard-spectator
contributions, involving the operators ${\cal O}_7$,
${\cal O}_8$ and penguin-type diagrams of ${\cal O}_1$. 
However, it has been stated that factorization holds,
at large recoil and leading order in $1/m_b$ [11]
and quantitative tests for proving QCD factorization
at the level of power corrections have been provided [17].

For a consistent treatment of radiative decays,
at next-to-leading order in QCD, a novel factorization
formula have been proposed in [12].
In this approach, the hadronic matrix elements in (1)
are written in terms of the essential form factor,
which describes the long-distance dynamics
and is a nonperturbative object, and
of the hard-scattering kernels, $T_i^I$ and $T_i^{II}$,
including the perturbative short-distance interactions,
as
\begin{equation}
\langle K^* \gamma | {\cal O}_i | \bar{B} \rangle
\, = \, \left[ F_{K^*} (0) T_i^I
+ \phi_B \otimes T_i^{II} \otimes \phi_{K^*}
\right] \cdot \eta \, ,
\end{equation}
where $\eta$ is the photon polarization.
When the dominant contribution comes from
${\cal O}_7$, we use (4) to write down
the decay amplitude as
\begin{eqnarray}
{\cal A}^{(0)} & = &
\frac{G_F}{\sqrt{2}} \,
\lambda_p \,
\frac{e m_b(\mu)}{2 \pi^2} \,
C_7 (\mu) \, F_{K^*}(0)
\nonumber \\*
& \times &
\left[ 
\varepsilon_{\mu \nu \alpha \beta}
\eta^{\mu} \epsilon^{* \nu} P_{K^*}^{\alpha} 
P_B^{\beta} 
- i \, (P_{K^*} q) (\eta \epsilon^* )
+ i  ( \epsilon^* q) (\eta P_{K^*} ) \right] ,
\end{eqnarray}
and consequently the branching ratio reads
\begin{equation}
Br^{LO} \, = \,
\tau_B \, \frac{G_F^2 \alpha |\lambda_p|^2 \,
m_b^2}{32 \pi^4} \; m_B^3 \, ( 1-z^2)^3 
| C_7 (m_b) |^2 |F_{K^*} (0)|^2 \; ,
\end{equation}
with $z=m_{K^*} /m_B$.
At next-to-leading order in $\alpha_s$,
one has to consider, in (6), the contributions to
the hard scattering kernels $T^I_i$
coming from the operators ${\cal O}_1$ and ${\cal O}_8$.
These have been evaluated in [12] and
bring (7) to the expression
\begin{eqnarray}
{\cal A} & = &
\frac{G_F}{\sqrt{2}} \,
\lambda_p \,
\frac{e m_b(\mu)}{2\pi^2} \left[
C_7 +
\frac{\alpha_s C_F}{4 \pi} \left(
C_1 G_1^p + C_8 G_8 \right) \right] F_{K^*}(0)
\nonumber \\*
& \times &
\left[ 
\varepsilon_{\mu \nu \alpha \beta}
\eta^{\mu} \epsilon^{* \nu} P_{K^*}^{\alpha} 
P_B^{\beta} 
- i \, (P_{K^*} q) (\eta \epsilon^* )
+ i  ( \epsilon^* q) (\eta P_{K^*} ) \right] ,
\end{eqnarray}
where $C_F = (N^2-1)/(2N)$, $N=3$, and
\begin{eqnarray}
G_1 (s) & = &
- \, \frac{833}{162} \, - \,
\frac{20 \, i \pi}{27} +
\frac{8 \pi^2}{9} \, s^{3/2}
\nonumber \\*
& & + \,
\frac{2}{9} \left[
48 + 30 i \pi - 5 \pi^2
- 2 i \pi^3 - 36 \zeta (3) + (36+6i \pi -9 \pi^2 )
\ln s \right.
\nonumber \\*
& & + \left. (3+6i \pi ) \ln^2 s + \ln^3 s \right] s
\nonumber \\*
& &
+ \, \frac{2}{9} \left[ 18 + 2 \pi^2 -2i \pi^3 +
(12 -6 \pi^2 ) \ln s + 6i \pi \ln^2 s +
\ln^3 s \right] s^2
\nonumber \\*
& & + \,
\frac{1}{27} \left[ -9 +112 i \pi -14 \pi^2
+ (182-48i \pi ) \ln s - 126 \ln^2 s \right] s^3 \, ,
\nonumber \\*
G_8 & = &
\frac{11}{3} - \frac{2 \pi^2}{9} + \frac{2i \pi}{3} \, ,
\end{eqnarray}
with $s_c = m_c^2/m_b^2$ and $\mu = m_b$.

Going further, we add factorizable
NLO hard-spectator corrections,
to the form factor $F_{K^*} (0)$.
For a single gluon exchanged with the
spectator (see Figure 1),
we extend the PQCD approach,
developed by Szczepaniak {\it et al.}
for {\it heavy-to-light} transitions dominated
by tree diagrams [15], to the
so-called {\it penguin} processes. \\
\marginpar{(Figure 1)}
\\ 
We evaluate the matrix element of the
operator ${\cal O}_7$,
as the following trace over spin, flavor and
color indices, and integration
over momentum fractions [16]
\begin{eqnarray}
T_{\mu} & = & {\rm Tr} \left[\bar{\phi}_{K^*}
\sigma_{\mu \nu} (1 + \gamma_{5}) q^{\nu} 
\frac{\rlap{/}{k}_{b} + m_{b}}{k_{b}^{2} -
m_{b}^{2}} \gamma_{\alpha} \phi_{B} \gamma^{\alpha} 
\frac{4 g_{s}^{2}}{Q^{2}} \right]   
\nonumber \\*
& + & \,  {\rm Tr} \left[\bar{\phi}_{K^*}
\gamma_{\alpha} \frac{\rlap{/} k_s}{k_{s}^{2}} 
\sigma_{\mu \nu}
(1 + \gamma_{5}) q^{\nu} \phi_{B} \gamma^{\alpha}
\frac{4 g_{s}^{2}}{Q^{2}} \right] ,
\end{eqnarray}
where $Q^2 \approx - (1-x)(1-y)m_B^2$.
The $B$ meson wave function
\begin{equation}
\phi_{B} = \frac{f_B}{12} \,
\varphi_{B}(x)(\rlap{/}{P}_{B} + m_{B})
\gamma_{5} 
\end{equation}
contains a strongly peaked distribution amplitude,
around $a =\lambda_B/m_B \approx 0.072$,
for $\lambda_B =0.38$.
The $K^*$ is described by the wave function
\begin{equation}
\phi_{K^*} = \frac{f_{K^*}}{12} \,
\varphi_{K^*}(y) \rlap{/}{P}_{K^*} \rlap{/}{\epsilon}
\, ,
\end{equation}
where the light-cone distribution amplitude,
$\varphi_{K^*}(y)$, has
the following expansion in Gegenbauer polynomials [18]
\begin{equation}
\varphi_{K^*} (y) = 6y(1-y)[ 1+ \alpha_1^{K^*} 
C^{(3/2)}_1(2y-1) + \alpha_2^{K^*} C^{3/2}_2(2y-1) + ...],
\end{equation}
with $C^{3/2}_1(u) = 3 u$,
$C^{3/2}_2(u) = (3/2)(5u^2-1)$,
$\alpha_1^{K^*} (m_b) = 0.18 \pm 0.05$, and
$\alpha_2^{K^*} (m_b) = 0.03 \pm 0.03$.
Performing the calculations in (11) and using
the form factors decomposition (4),
we identify the spectator contribution to the
essential form factor as
\begin{equation}
F^{sp} (a) \, = \, \frac{g_{s}^{2}}{9}
\frac{f_B f_{K^*}}{m_B \lambda_B } 
\int_0^{1-a} dy \; \frac{(2-y)}{(1-y)^2} \,
\varphi_{K^*} (y) \, ,
\end{equation}
where the $K^*$-mass is neglected.
Since we have introduced a cut-off for $y \to 1$,
the form factor correction (15)
depends on the peaking parameter $a$
and this is a main uncertainty in our calculations.
For the following input values:
$\alpha_s (\mu = Q^2) \approx 0.38$, $f_B = 0.180$ GeV,
$f_{K^*}^{\perp} = 0.185$ GeV
and $a=0.072$, we get
\begin{equation}
F^{sp} (0.072) \, = \, 0.1475
\end{equation}
With the total form factor
$F_{K^*}(0) + F^{sp} (0.072) =
0.38 + 0.1475$ in the amplitude (9),
the branching ratio gets significantly enhanced
to the value $Br^{NLO} = 6.97 \times 10^{-5}$.
This is comparable to the average theoretical
prediction $\left( 7.5 \pm 0.3 \right)
\times 10^{-5}$  [9, 11, 12],
but is much above the experimental data:
\begin{eqnarray}
Br(B^+ \to K^{*+} \gamma) =
\left \lbrace
\begin{array}{lc}
\left( 3.83 \pm 0.62 \pm 0.22 \right) \times 10^{-5}
& ({\rm BaBar} \; [3]) \\
\left(  3.76_{-0.83}^{+0.89} \pm 0.28
\right) \times 10^{-5} & ( {\rm CLEO} \; [4]) \\
\left( 3.89 \pm 0.93 \pm 0.41 \right)
\times 10^{-5} & ({\rm Belle} \; [5])
\end{array}
\right.
\nonumber \\*
Br(B^0 \to K^{*0} \gamma) =
\left \lbrace
\begin{array}{lc}
\left( 4.23 \pm 0.40 \pm 0.22 \right) \times 10^{-5}
& ({\rm BaBar} \; [3]) \\
\left( 4.55^{+0.72}_{-0.68} \pm 0.34 
\right) \times 10^{-5} & ({\rm CLEO} [4]) \\
\left( 4.96 \pm 0.67 \pm 0.45 \right) \times
10^{-5} & ({\rm Belle} [5])
\end{array}
\right.
\nonumber
\end{eqnarray}

Moreover, the direct CP asymmetry, defined as
\begin{equation}
a_{CP} \, = \, \frac{\Gamma(\bar{B} \to K^* \gamma ) -
\Gamma ( B \to K^* \gamma )}{
\Gamma(\bar{B} \to K^* \gamma ) +
\Gamma ( B \to K^* \gamma )} \, ,
\end{equation}
has been predicted by the SM to
be $a_{CP} < | 0.005 |$, and this disagrees
with the BaBar and CLEO data,
$a_{CP} = - 0.044 \pm 0.076 \pm 0.012$ (BaBar [3])
and $a_{CP} = 0.08 \pm 0.13 \pm 0.03$ (CLEO [4],
for the sum of neutral and charged $B \to K^* \gamma$ decays).

So, predictions for measurable values of the direct
CP asymmetry, in agreement with more precise measurements,
might have a dominantly new physics origin.
Following this idea, let us analyse
the $B \to K^* \gamma$ decay in the MSSM context.
Using the mass insertion approximation, [19],
we incorporate, in the Wilson coefficients
$C_{7}$ and $C_8$, the FCNC SUSY contributions 
\begin{eqnarray}
C_{7}^{SUSY} (M_{SUSY})  & = & 
\frac{\sqrt{2} \pi \alpha_s}{G_F (V_{ub} V_{us}^* +
V_{cb} V_{cs}^*) m_{\tilde{g}}^2} \, 
\left( \delta_{23}^d \right)_{LR} 
\frac{m_{\tilde{g}}}{m_b} \, F_0(x) \, ; 
\nonumber \\*
C_{8}^{SUSY} (M_{SUSY}) & = & \frac{\sqrt{2} \pi
\alpha_s}{G_F (V_{ub} V_{us}^* +
V_{cb} V_{cs}^*) m_{\tilde{g}}^2} 
\left( \delta_{23}^d \right)_{LR}
\frac{m_{\tilde{g}}}{m_b} \, G_0(x) \, ,
\end{eqnarray}
where
\begin{eqnarray}
F_0 (x) & = & - \;
\frac{4x}{9(1-x)^4} \,
\left[ 1+4x-5x^2+4 x \ln(x) + 2 x^2 \ln (x) \right] ,
\nonumber \\*
G_0 (x) & = &
\frac{x}{3(1-x)^4} \,
\left[ 22-20x-2x^2+16 x \ln(x) -x^2 \ln (x)
+ 9 \ln (x) \right] \; \; \; \;
\end{eqnarray}
In (19), $x=m_{\tilde{g}}^2 / m_{\tilde{q}}^2$
is expressed in terms of the gluino mass, $m_{\tilde{g}}$,
and an average squark mass, $m_{\tilde{q}}$.
We underline that, in the expressions of
$C_{7,8}^{SUSY} (M_{SUSY})$,
we have kept only the left-right squark mixing parameter 
$\left( \delta_{23}^d \right)_{LR} \, = \,
\left( \Delta_{bs} \right)/
m_{\tilde{q}}^2$
since,
being proportional to the large
factor $m_{\tilde{g}} /m_b$, will have a significant
numerical impact on the branching ratio value.
The quantities $\Delta_{bs}$ are the
off-diagonal terms in the sfermion mass matrices,
connecting the flavours $b$ and $s$ along the sfermion
propagators [19]. In these assumptions,
the total Wilson coefficients, encoding the New Physics, become
\begin{eqnarray}
C_{7}^{total} [x, \delta] & = &
C_{7} (m_b) + C_{7}^{SUSY} (m_b) \; ,
\nonumber \\*
C_{8}^{total} [x, \delta] & = &
C_{8} (m_b) + C_{8}^{SUSY} (m_b) \; , 
\end{eqnarray}
where
$C^{SUSY}_{7,8} (m_b)$ have been evolved from
$M_{SUSY} = m_{\tilde{g}}$ down to the
$\mu =m_b$ scale, using the relations [20]
\begin{eqnarray}
C_{8}^{SUSY} (m_b) & = &
\eta C_{8}^{SUSY}(m_{\tilde{g}} ) \, ,
\nonumber \\
C_{7}^{SUSY} (m_b) & = & \eta^2
C_{7}^{SUSY}(m_{\tilde{g}} ) + \frac{8}{3}
(\eta - \eta^2) C_{8}^{SUSY}(m_{\tilde{g}} ) \, ,
\end{eqnarray}
with
\begin{equation}
\eta \, = \,
\left( \alpha_s(m_{\tilde{g}})/
\alpha_s(m_t) \right)^{2/21}
\left( \alpha_s(m_t)/
\alpha_s(m_b) \right)^{2/23}
\end{equation}

Finally, putting everything together, we replace,
in (9), the Wilson coefficients $C_7$ and $C_8$
respectively by $C_7^{total}[x , \delta]$ and
$C_8^{total} [x , \delta ]$, the form factor
$F^{K^*} (0)$ by $F^{K^*} (0) + F^{sp} (a)$
and, consequently, the branching ratio (8) turns into
\begin{eqnarray}
Br^{total} & = & Br^{SM + SUSY} \, = \,
\tau_B \, \frac{G_F^2 \alpha m_b^2}{32 \pi^4} \,
m_B^3 \, ( 1-z^2)^3 \; 
|F_{K^*} (0) +F^{sp} (a)|^2
\nonumber \\*
& \times &
\left| \lambda_p \left[ C_7^{total} [x, \delta ] + 
\frac{\alpha_s C_F}{4 \pi} \left(
C_1 G_1^p + C_8^{total} [x, \delta ] 
G_8 \right) \right] \right|^2
\end{eqnarray}
One can notice that, for a given $x$ and
$\delta \equiv \rho e^{i \varphi}$,
the total branching ratio
is depending on three free parameters:
$a , \, \rho , \, \varphi$, while the direct
CP asymmetry parameter, (17),
is free of the uncertainty $a$.

In the next coming discussion, we use the following
input parameters: $m_b (m_b)=4.2$ GeV,
$\alpha =1/137$, $|V_{tb}V_{ts}^*| = 0.0396 \pm 0.002$,
$\tau_{B^0} = \left( 1.546 \pm 0.018 \right)$ ps,
the QCD sum rules analyses result
$F_{K^*} (0) = 0.38$, and $m_{\tilde{q}} = 500$ GeV.
In what it concerns the gluino, as its pair production
cross section has large cancellations
in the $e^+ e^-$ annihilation, there is hope
that the laser-backscattering photons will provide
a precise gluino mass determination [21].
For a wide range of squark masses,
a gluino mass of 540 GeV may be measured,
with a precision of at least
$\pm 2 \dots 5$,
at the multi-TeV linear collider at CERN.\\
\marginpar{(Figure2)}\\

For $x$ taking the values $x_l =0.3$, 
$x_0 =(540/500)^2$ and $x_g =3$
(where $l(g)$ comes from $m_{\tilde{g}}$ less (greater)
than  $m_{\tilde{q}}$),
and imposing the BaBar constraint [3]
\begin{equation}
- \, 0.17 < a_{CP} < 0.082 \, ,
\end{equation}
we draw, in Figure 2, the contour plots
of constant $Br^{total}$ (the dashed lines) and
$a_{CP}$ (the solid lines).
When $\left \lbrace \rho , \, \varphi
\right \rbrace \in [0, \, 0.03] \times 
[ -  \pi /2 , \, \pi /2 ]$, we get,
for the world average branching ratio data,
over the $B^{\pm}$ and $B^0$ decay modes,
$Br_{exp} (B^{\pm} \to K^{*\pm} \gamma )
\, = \, \left( 4.22 \pm 0.28 \right) \times 10^{-5}$,
three dashed lines, with increasing thickness,
as $x$ goes from $x_l$ to $x_g$.
Correspondingly, for $a_{CP}$,
we get three pairs of solid curves:
the lower ones, for $a_{CP} = - 0.17$,
and the upper ones, for $a_{CP} =0.082$.
These solid contours close inside
the values of direct CP asymmetry which
do not agree with (24).
Now, we are able to put strong
constraints on $\left( \delta_{23}^d \right)_{LR}$,
by looking at the segments of the $Br$-plots
outside the solid contours, for each $x$.
We notice that, for $m_{\tilde{g}} > m_{\tilde{q}}$,
all the negative phases, with suitable
$\rho$'s, can accommodate both the relation
(24) and the branching ratio data. \\
\marginpar{(Figure 3)}\\

In Figure 3, we represent, the $Br^{total}$
(in units of $10^{-5}$) and $a_{CP}$
(in units of $0.1$), with respectively
dashed and solid lines, as functions of $\varphi$,
for $x=3$.
As $\rho$ takes the following values:
$\rho \in \left \lbrace 0.005 , \, 0.01 , \,
0.015 , \, 0.02 \right \rbrace$, we get four pairs
of curves, with increasing thickness.
The horizontal dashed line corresponds to the
average $Br$ data, while the horizontal solid ones
stand for the constraint (24).

Finally, let us perform a numerical analyses,
for $x=x_0$, and increasing $\rho$,
starting with $\rho =0.005$.
As $\varphi \in \left[ - 8 \pi/16 , \, - 4 \pi/16 
\right] \cup \left[ 3 \pi/16 , \, 7 \pi/16 \right]$,
the $Br^{total}$ and the direct CP asymmetry
are inside the ranges
$10^5 \times Br^{total} \in 
\left[ 8.3 , \, 3.3 \right]$
and $\left[ 3.1 , \, 8.1 \right]$ and, respectively,
$a_{CP} \in \left[ -0.054 , \, - 0.093 \right]
\cup \left[ 0.096 , \, 0.062 \right]$,
accommodating data and other theoretical
models predictions.
When $\rho$ goes to bigger values,
the two $\varphi$ ranges, constrained by
the allowed branching ratios, get closer and
$a_{CP}$ moves toward much bigger values.
For example, for $\rho = 0.01$ and
$\varphi \in \left[ - 6 \pi/16 , \, - 4 \pi/16 \right] 
\cup \left[ 3 \pi/16 , \, 5 \pi/16 \right]$,
one gets $10^5 \times Br^{total} 
\in \left[ 8.1 , \, 3.6 \right]$ and
$\left[ 3.2 , \, 7.62 \right]$ and, respectively,
$a_{CP} \in \left[ -0.1 , \, - 0.16 \right]
\cup \left[ 0.21 , \, 0.12 \right]$ and
we notice that only the negative $\varphi$-values
lead to $a_{CP}$ inside the BaBar constraint (24).
For $\rho = 0.015$ and
$\varphi \in \left[ - 4 \pi/16 , \, - 2 \pi/16 \right] 
\cup \left[ \pi/16 , \, 3 \pi/16 \right]$,
the values $10^5 \times Br^{total}
\in \left[ 7.8 , \, 3.6 \right]$
and $\left[ 3.3 , \, 7.3 \right]$
are compatible with a measurable
$a_{CP} \approx \pm 0.12$.
Starting with $\rho =0.02$, the predictions for
branching ratio
are above data and other theoretical estimations,
the minimum value being
$Br^{total} = 9 \times 10^{-5}$, for $\varphi =0$,
while the corresponding asymmetry is $a_{CP} = 0.003$.

In the present paper, we have analysed
the radiative $B \to K^* \gamma$ decay, in a combined
PQCD and SUSY framework.
First, we have used the PQCD approach, developed
by Szczepaniak {\it et al.} [15] and extended to
^^ ^^ penguin'' processes [16], to compute
the hard-spectator contribution, $F^{sp} (a)$, to the
essential form factor $F_{K^*} (0)$.
For the peaking parameter in the $B$ wave function
$a=0.072$ and $F_{K^*} (0) =0.038$,
the branching ratio becomes
$Br^{NLO} = 6.97 \times 10^{-5}$, which is
above the experimental data,
while the direct CP asymmetry
predicted by the SM lies much below [3, 4].
In order to find an agreement,
we extend our analyses by
including, in the Wilson coefficients $C_{7,8}$,
the SUSY contributions coming
from squark mixing parameter
$\left( \delta_{23}^d \right)_{LR} = \rho e^{i \varphi}$.
Consequently, the total branching ratio depends,
besides $a$, on three (SUSY)
parameters: $x , \rho , \varphi$, while $a_{CP}$
is free of the uncertainty coming from the form factors.
Using the graphs displayed in Figures 2 and 3,
one is able to find out allowed ranges for
the mass insertion parameter
$\left( \delta_{23}^d \right)_{LR}$.
As an example, for $x= (540/500)^2$,
the world average branching ratio, $Br_{exp} =
4.22 \times 10^{-5}$, can be accommodated for
$\lbrace \rho , \, \varphi \rbrace
= \left \lbrace 0.005 , \, - \, \frac{4 \pi}{13}
\right \rbrace$ or
$\left \lbrace 0.01 , \, - \, \frac{4 \pi}{15}
\right \rbrace$.
The corresponding asymmetries, $a_{CP} =-0.085$ 
and respectively $a_{CP} = -0.147$, are inside the
BaBar constraint (24).

\begin{flushleft}
{\bf ACKNOWLEDGMENTS}
\end{flushleft}
The authors gratefully acknowledge the
kind hospitality and fertile environment
of the University of Oregon
where this work has been carried out.
Professor N.G. Deshpande's clarifying discussions
and constant support are highly regarded.
M.A.D.'s gratitude goes to
the U.S. Department of State, the Council
for International Exchange of Scholars (C.I.E.S.)
and the Romanian-U.S. Fulbright Commission, for
sponsoring her participation in the
Exchange Visitor Program no. G-1-0005.

\newpage

\begin{center}
{\bf FIGURE CAPTIONS}
\end{center}
Fig.1. The Feynman contributing diagrams in the
hard scattering amplitude $T_{\mu}$. The gluon and
photon are respectively represented
by dotted and dashed lines.\\
Fig.2. Contour plots of total branching ratio
fitting the world average data,
in units of $10^{-5}$, (the dashed lines)
and the BaBar constraint (24) on direct CP asymmetry
(the solid lines), as functions of $\rho$ and
$\varphi$. The thickness of contours is increasing
as $x$ takes the values:
$x = \left \lbrace 0.3 , \, 1.16 , \, 3 \right \rbrace$.
The solid curves close inside the values of
$a_{CP}$ which disagree with the constraint (24). \\
Fig.3. The total branching ratio, in units of
$10^{-5}$, (the dashed lines) and $10 \times a_{CP}$
(the solid lines), as functions of $\varphi$, for $x=3$.
The thickness of plots increases
as $\rho$ is respectively:
$\rho = \lbrace 0.005 , \, 0.01 , \,
0.015 , \, 0.02 \rbrace$.
The horizontal dashed line corresponds to the
world average branching ratio and the
horizontal solid lines are for the constraint (24).

\end{document}